# DICKE COHERENT NARROWING IN TWO-PHOTON AND RAMAN SPECTROSCOPY OF THIN VAPOUR CELLS

Gabriel DUTIER, Petko TODOROV, Ismahène HAMDI, Isabelle MAURIN,

Solomon SALTIEL, Daniel BLOCH and Martial DUCLOY

*Laboratoire de Physique des Lasers, UMR7538 du CNRS et de l'Université Paris13*

*99 Av JB Clément, F 93430 Villetaneuse,*   mail *: bloch@lpl.univ-paris13.fr*

Abstract

The principle of coherent Dicke narrowing in a thin vapour cell, in which sub-Doppler spectral lineshapes are observed under a normal irradiation for a $\lambda/2$ thickness, is generalized to two-photon spectroscopy. Only the sum of the two wave vectors must be normal to the cell, making the two-photon scheme highly versatile. A comparison is provided between the Dicke narrowing with copropagating fields, and the residual Doppler-broadening occurring with counterpropagating geometries. The experimental feasibility is discussed on the basis of a first observation of a two-photon resonance in a 300 nm-thick Cs cell. Extension to the Raman situation is finally considered.

Linear spectroscopy in a macroscopic low-pressure gas is usually sensitive to the Doppler broadening, as governed by $|\vec{k}|u$ (with u the most probable thermal velocity, and $\vec{k}$ the wave vector of the irradiating field). In c.w. multiphoton spectroscopy, when several irradiating beams ($\omega_i$, $\vec{k}_i$) are used, the Doppler broadening generalizes to $|\sum \vec{k}_i|u$ [1]. When a cell of dilute gas is so thin that atomic (or molecular) trajectories are mostly from wall-to-wall, sub-Doppler spectral lineshapes can be obtained in a variety of processes (optical pumping, ...) under *normal* incidence : this results from a specific enhancement of the slow atoms contribution (see [2] and refs. therein), with respect to the transient nature of the atom-light interaction. In linear optical absorption, a particular coherent narrowing has been experimentally observed for a $\lambda/2$ cell thickness [3,4], analogous to the one first mentioned by Romer and Dicke [5] in the microwave domain. The purpose of this note is to show that the main features of this coherent Dicke narrowing can be extended to two- (or multi-) photon transitions, as long as an effective two-level model can be considered, and for a *normal* incidence condition now referring to the sum of photon wave-vectors $|\sum \vec{k}_i|$.

Let us first recall that for a cell filled with two-level atoms, the 1-photon resonant transmission (or reflection [6-7]) behaviour depends on the field radiated by the induced macroscopic polarization, resulting from the spatial summing of the local optical coherence, as given by the quantity Re[$I_f$] with:

$$I_f = i \int_0^L \sigma_{21}(z)dz = i\int_{-\infty}^{+\infty}\int_0^L \sigma_{21}(z,v_z)f(v_z)dv_z dz \qquad (1)$$

In (1), L is the cell length, Oz is the axis normal to the thin cell, $f(v_z)$ the velocity distribution, usually assumed to be Maxwellian. Assuming a linear regime of interaction, $\sigma_{21}(z,v_z)$ is given as a result of the elapsed time $t = z/v_z$ from the wall departure (for $v_z > 0$), where $\sigma_{21}=0$, by :

$$\sigma_{21}(z,v_z) = \frac{i\Omega}{2D_{21}}[1-\exp(-D_{21}z/v_z)] \qquad (2)$$

In (2), $\Omega$ is the Rabi frequency, and $D_{21} = \gamma_{21} - i(\Delta - \vec{k}.\vec{v})$ characterizes the resonance with $\gamma_{21}$ the relaxation coefficient of the optical dipole and $\Delta$ the frequency detuning between the field and the $|1\rangle \rightarrow |2\rangle$ transition. Note that assuming a normal incidence, the Doppler shift has been counted on the same direction as the one along which the interaction time is measured (*i.e.* for $\vec{k}.\vec{v} = kv_z$). For $v_z < 0$, $z/v_z$ is replaced by $(z - L)/v_z$. The transmission lineshape strongly depends on the L thickness [3, 4, 8]: it oscillates, with a pseudo-periodicity $\lambda$, from a sub-Doppler width (for $L = (2n+1)\lambda/2$) to a Doppler-broadened one (for $L = n\lambda$). The narrowest lineshape is for the $\lambda/2$ thickness as in the situation analyzed by Romer and Dicke [5]. It exhibits a logarithmic singularity (in the limiting case $ku \gg \gamma_{21}$), originating in the $1/v_z$ contribution of fast atoms. Also, in spite of a mismatching brought by the Doppler detuning [3,4], all velocity groups interfere constructively up to the $\lambda/2$ thickness.

In the extension to a two-photon situation -see fig. 1-, one monitors the $(\omega_2, \vec{k}_2)$ probe beam transmission change induced by a $(\omega_1, \vec{k}_1)$ pump, governed by an integrated optical coherence $I_f^{2-ph} = i\int_0^L \sigma_{32}(z)dz$. In an effective two-level model situation, *i.e.* when the irradiating frequencies $\omega_1$ and $\omega_2$ are far enough from the resonances $\omega_{21}$ and $\omega_{32}$, the pump-induced optical coherence $\sigma_{32}(z)$ simply results from the transient regime of the resonant two-photon coherence $\sigma_{31}(z)$. The assumption is that the sum of the exciting frequencies $\omega_1+\omega_2$ is nearly two-photon resonant, *i.e* that $\Delta_{31} = (\omega_1+\omega_2)-\omega_{31}$ remains comparable with the two-photon width (homogenous width $\gamma_{31}$, or at least with the Doppler broadened width $|\vec{k}_1 + \vec{k}_2|u$), while the individual excitation frequencies $\omega_1$, $\omega_2$ are largely detuned from the 1-photon resonances $\omega_{21}$ and $\omega_{32}$. In this situation, saturation effects are most often negligible and a third order perturbation expansion (second order with respect to the "pump" field, first order with respect to the detected "probe" field) is sufficient to evaluate the atomic response $\sigma_{32}$.

The essence of the extension of the coherent Dicke narrowing to two-photon transition lies in the fact that the $\sigma_{32}(z, v_z)$ behaviour is strictly analogous to the one of $\sigma_{21}(z, v_z)$ in (1), with the 1-photon $\vec{k}$ vector simply replaced by the 2-photon wave-vector $\vec{k}_1 + \vec{k}_2$. Namely, one shows (for $v_z > 0$) that :

$$\sigma_{32}(z, v_z) \approx \frac{i\Omega_1^2 \Omega_2}{8\Delta^2 D_{31}} \left[1 - \exp(-D_{31} z / v_z)\right] \quad (3)$$

(and $z/v_z$ replaced by $(z - L)/v_z$ for $v_z < 0$ ). In (3), $D_{31} = \gamma_{31} - i(\Delta_{31} - (\vec{k}_1 + \vec{k}_2) \cdot \vec{v})$ characterizes the 2-photon resonance, while $\Delta$ is the detuning between the pump laser frequency and the 1-photon excitation, and $\Omega_1$ and $\Omega_2$ are the respective Rabi frequencies associated to the pump ($\omega_1$) and to the probe ($\omega_2$) beams. In the vicinity of the two-photon resonance, the large value of $\Delta$ implies that $\Delta = \omega_1 - \omega_{21} \approx \omega_{32} - \omega_2 \equiv \Delta_2$.

A justification for (3) can be obtained from an exact calculation of $\sigma_{32}$, as provided in the appendix A in the non-restrictive frame of a conservative 3-level system. The full analogy of (3) with (2) is responsible for the two-photon coherent Dicke narrowing, but it should be noted that the complex phase factor now depends upon the *vectorial sum of the wavevectors* $\vec{k}_1 + \vec{k}_2$. As a consequence, the width evolves from a Doppler broadening $|\vec{k}_1 + \vec{k}_2| u$ to a sub-Doppler structure depending on $\gamma_{13}$, and the maximal narrowing is hence obtained for a thickness $L = \Lambda/2$ with $\Lambda = 2\pi/|\vec{k}_1 + \vec{k}_2|$. Also, in such a two-photon extension, nontrivial geometries are allowed, such as an oblique incidence for the irradiating beams, with $\vec{k}_1 + \vec{k}_2$ remaining along the normal. This versatility may be used as an advantage by choosing an irradiation close to the Brewster angle (provided that the wavelengths of pump and probe are not too different) (see fig. 2, and [9]), hence eliminating the Fabry-Perot effects [7] associated with ETC spectroscopy. Also, it opens the possibility to optimize, with the choice of incidence angles, the Dicke narrowing for a given cell thickness.

The above description allows for a comparison between the co- and counter-propagating geometries. When the two irradiating frequencies are nearly equal ($\omega_1 \approx \omega_2$), a counter-propagating geometry provides a negligible Doppler broadening in a macroscopic cell, while the co-propagating geometry is affected by a Doppler broadening twice as large as the one for 1-photon transition. In contrast, for an ETC with a thickness suitable for the 2-photon Dicke narrowing in the co-propagating geometry -i.e. $L = \Lambda/2 = \lambda_1.\lambda_2/2(\lambda_1+\lambda_2)$ ($\approx \lambda_1/4 \approx \lambda_2/4$)- , the counter-propagating geometry corresponds to an ETC thickness much smaller than the optimal one (*i.e.* $L<<\Lambda$ with $\Lambda = \lambda_1 \lambda_2 /|\lambda_1 - \lambda_2|$), implying a spectral regime dominated by transit time effects, so that the two-geometries provide comparable spectra (see fig. 3). Moreover, when $\omega_1$ and $\omega_2$ largely differ, the sub-Doppler Dicke narrowing can still be attained (in the co-propagating geometry for a specific well-chosen thickness, or if preferable, with an adapted oblique geometry), while the counter-propagating geometry, that would be sensitive to a partial Doppler broadening in a macroscopic cell, actually undergoes a reduction of this Doppler broadening through transit time effects, at the expense of a drastic reduction of the contributing velocities. Besides, a frequency-modulation (FM) technique [3] can turn these narrow spectra into Doppler-free lineshapes.

The possible observation, and benefits, of such a 2-photon coherent narrowing could be limited by the large 1-photon detuning $\Delta$, with respect to a signal decreasing with $1/\Delta^2$. Because tuneable sources open the possibility to strongly increase the strength of a two-photon signal by coming close to a nearly-resonant 1-photon excitation, it is worth discussing practical limitations. It is essential for the validity of (3) that $\Delta$ (or $D_{21}$ and $D_{32}$, see appendix A) largely exceeds $|v_z|/L$. In spite of the fact that the 2-photon Dicke narrowing (for $L = (2n+1)\Lambda/2$) induces a sub-Doppler structure through an overweighed contribution of the slow atoms, the coherent (and constructive) contribution of atoms with thermal velocities cannot be neglected, even at line-centre, and the requirement: exp-$[\Delta\Lambda/2v_z] \ll 1$ applies also to fast (*i.e.* thermal) atoms, practically implying for the detuning to be in excess of several Doppler widths. However, for an oblique and large angle incidence (*e.g.* fig. 2), the 2-photon Dicke-type narrowing could remain observable, even for a relatively weak detuning $\Delta$: indeed, the successive 1-photon processes cannot eliminate the Doppler-broadening when two different $\vec{k}_1$ and $\vec{k}_2$ axes are involved, so that the stepwise excitation process cannot strongly interfere with the 2-photon process. As a counterpart of the large pump detuning, the pump intensity can be high, as the condition $\Omega_1, \Omega_2 \ll |\Delta|$ required by the 3$^{rd}$ order perturbation expansion, that imposes a Stark shift much smaller than $\Delta$, remains easily respected.

With respect to these sensitivity limitations, it is worth reporting a previous observation, although performed in a counter-propagating geometry, of a direct two-photon transition in a sub-micrometric cell. In the context of ETC spectroscopy on the stepwise $6S_{1/2}$ -($6P_{3/2}$)- $6D_{5/2}$ transition of Cs, the direct two-photon excitation was observed in an extended frequency scan. The pump frequency $\omega_1$, locked onto the 1-photon 852 nm resonance for the $6S_{1/2}$ (F=3) $\rightarrow$ $6P_{3/2}$ transition, is indeed a blue-detuned pump for the $6S_{1/2}$ (F=4) $\rightarrow$ $6P_{3/2}$ transition, with the $\Delta$ = 9.192 GHz detuning largely exceeds any relevant width for a 2-photon transition. This justifies that when scanning the 917 nm probe on the $6P_{3/2}$-$6D_{5/2}$ transition, one observes (fig. 4), aside from the stepwise 2-photon resonance, a red-shifted resonance associated to the direct two-photon transition $6S_{1/2}$ (F=4) $\rightarrow$ $6D_{5/2}$.

The main information derived from this recording is that, in spite of the unusually small size (L = 315nm) of the cell, the two-photon signal is observable, and here of a relatively large size. The absorption is in excess of $10^{-3}$, and the signal-to-noise ratio is easily improvable, as for the specific purpose of the stepwise excitation experiment, the sensitivity of the set-up was non-optimized and limited to ~$10^{-4}$. Also, the two-photon signal, observed here for a moderate pump intensity (1mW focused on a ~100µm spot, *i.e.* $\Omega_1$ ~ 250MHz), could be largely enhanced by decreasing the 9.192 GHz pump detuning, while remaining in the frame of a non resonant 2-photon transition. These potential improvements should enable the observation of the two-photon signal at a lower atomic density, the effect of which seems to be the dominant reason for the unexpectedly broad two-photon signal (the $6D_{5/2}$ hyperfine structure accounts only for ~50MHz, the counterpropagating geometry only allows a ~20 MHz 2-photon Doppler broadening, and the transit time is typically in excess of 1.5ns - *i.e.* a broadening below 100 MHz -). It can be also noted that the direct 2-photon signal is less sensitive to the pressure broadening than the stepwise excitation because resonance collision processes affect the intermediate transitions.

In conclusion, we have demonstrated that the recently demonstrated coherent Dicke narrowing in an ETC can be extended through versatile set-ups to 2-photon transitions, and a genuine experimental observation appears quite feasible. Even if the Dicke narrowing may remain only of a marginal interest for precision spectroscopy (the fabrication of ETCs [4,10] requires thick windows for a small active thickness), several possible applications can be envisioned. ETCs are already an interesting tool for the probing of atom-surface interaction, notably for the effective analysis of the spatial dependence of long-distance atom-surface interaction [11]. The high-lying excited states, of a special interest in this context, can be conveniently reached with 2- or multi- photon excitation. Direct transitions from the ground states are naturally simpler to analyze than multiple stepwise excitation : in particular there is no need to assume a quasi-thermal pumping (see *e.g.*[12]), hardly justified in an ETC, and that ignores the surface interaction that shifts the 1-photon resonance. Moreover, because there is a physical upper limit to the atom-surface distance, the effect of atom-surface interaction in a multi-photon transition should be simpler to analyze in ETC spectroscopy than in the alternate selective reflection (SR) spectroscopy (see [12]) : in this last case, the effective length of the probed region under multiple beam irradiation has never been determined unambiguously in spite of various principle analysis [13]. Extension of the multi-photon Dicke narrowing to Λ-type systems based upon Raman transitions in a counter-propagating geometry can also be predicted, with $\vec{k}_1 - \vec{k}_2$ the relevant wavevector. For Raman transitions between sublevels of the ground state, the expected long lifetime of the Raman coherence should make observable a large number of collapses and revivals of the Dicke narrowing (for Cs, and a ground state coherence lifetime assumed to be ~1 μs, the decoherence length is ~ 200 μm, to be compared with a Dicke periodicity Λ=426 nm for an excitation close to the $D_2$ line with counter-propagating beams). Hence, with relatively long cells (~10-100 μm), a new-type of ground state Raman coherence analysis appears feasible. The effective decay of the Raman coherence narrowing, measured through an adjustable cell length or through an angular adjustement of the respective exciting beams, could yield sensitive information on how the environment -and confinement- affects the details of the relaxation of the Raman coherence.

*This work has been partially supported by FASTNet (European contract HPRN-CT-2002-00304) and by the RILA (#09813UK ) French-Bulgarian cooperation. We acknowledge the participation of L. Petrov to some of the experimental steps.*

**Appendix A**

We give below the spatial dependence of the transient optical coherence $\sigma_{32}$. The calculation is obtained in the frame of a third-order perturbation expansion, and for the sake of simplicity, a conservative atomic system is assumed, with negligible dephasing collisions. In spite of the non resonant nature of the pump excitation, the rotating wave approximation is performed also with respect to the pump, simply assuming that $|\omega_1 - \omega_{12}| \ll |\omega_1 + \omega_{12}|$ : our scope is indeed to define conditions for which the elementary two-photon processes dominates in the competition with the nearly resonant stepwise (1-photon) processes. Under such hypotheses, one gets (for $v_z > 0$):

$$\sigma_{32}(z,v) = \frac{i\Omega_1^2 \Omega_2}{8} \left[ \begin{array}{l} \dfrac{[1-\exp-(D_{31}z/v)]}{D_{31}(D_{21}-D_{31})(D_{32}-D_{31})} + \dfrac{[1-\exp-(D_{21}z/v)]}{D_{21}(D_{21}-D_{32})(D_{21}-D_{31})} + \dfrac{[1-\exp-(D_{32}z/v)]}{D_{32}(D_{32}-D_{21})(D_{32}-D_{31})} + \\ + \dfrac{[1-\exp-(D_{31}z/v)]}{D_{31}|D_{21}|^2}\exp(-D_{21}^* z/v) + \dfrac{[1-\exp-(D_{32}z/v)]}{D_{32}|D_{21}|^2}[1-\exp(-2\gamma_{21}z/v)] \end{array} \right]$$

(A.1)

with $D_{32} = \gamma_{32} + ik_2 v - i\Delta_2$ ; $\gamma_2 = 2\gamma_{21}$, $\gamma_{31} = \gamma_3/2$ and $\gamma_{32} = (\gamma_3 + \gamma_2)/2 = \gamma_{31} + \gamma_{21}$. Note that because of the absence of dephasing collisions, one has $D_{31} = D_{32} - D_{21}^*$. In (A.1), all terms but two exhibit a denominator on the order of $\Delta^3$ (assuming that the 1-photon detuning $\Delta$ largely dominates over all other broadenings). Among the two terms with a resonant denominator on the order of $D_{31}\Delta^2$, the first one is simply the 2-photon coherence contribution appearing in (3) (after some firther simplifications). Surprisingly enough, the other one originates in a transient build-up of population in the relay level $|2\rangle$, but the exponential factor shows that the contribution associated with the build-up of population induced in the level $|2\rangle$ is limited to a very short duration $\sim 1/\Delta_2$ ($\approx 1/\Delta$), during which the system does not have time to experience the 1-photon detuning. Hence, for any reasonable length of the ETC, only the resonant two-photon coherence survives.

**Figure Captions**

Figure 1 : A schematic of the atomic 3-level system

Figure2 : A non-collinear geometry, close to the Brewster angle $i_B$, enabling to observe the two-photon Dicke narrowing when ($\vec{k}_1 + \vec{k}_2$) is perpendicular to the thin cell.

Figure 3 : A comparison between the co-propagating geometry (full line), and the counterpropagating geometry (dashed line). The conditions are $\omega_{12}=\omega_{23}$, $\lambda_1 \approx \lambda_2$, $L = \lambda_1/4 = \lambda_2/4$, enabling strict Doppler-free 2-photon spectroscopy in a macroscopic cell. One has taken $\gamma_{31} = k_2 u/40$. The calculation is independent of the detuning $\Delta$.

Figure 4 : Transmission spectrum recorded across the excited transition $6P_{3/2}$-$6D_{5/2}$ (917 nm) in a counterpropagating geometry on a 315 nm-thick Cs cell (T= 220°C) with the pump beam locked onto the $6S_{1/2}$(F=3)-$6P_{3/2}$ transition. The left spectrum is a zoom on the auxiliary resonance corresponding to the 2-photon resonance $6S_{1/2}$(F=4)-$6D_{5/2}$. The pump frequency is locked to the $6S_{1/2}$(F=3)-$6P_{3/2}$ transition. The hyperfine structure of the $6S_{1/2}$-$6P_{3/2}$-$6D_{5/2}$ levels is also shown.

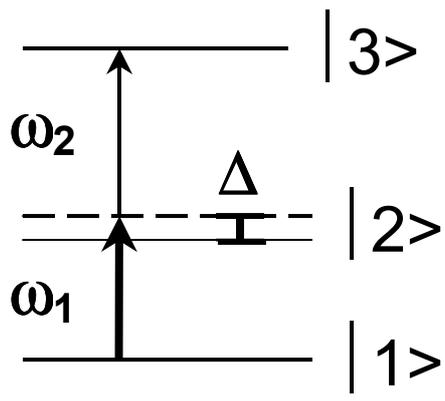

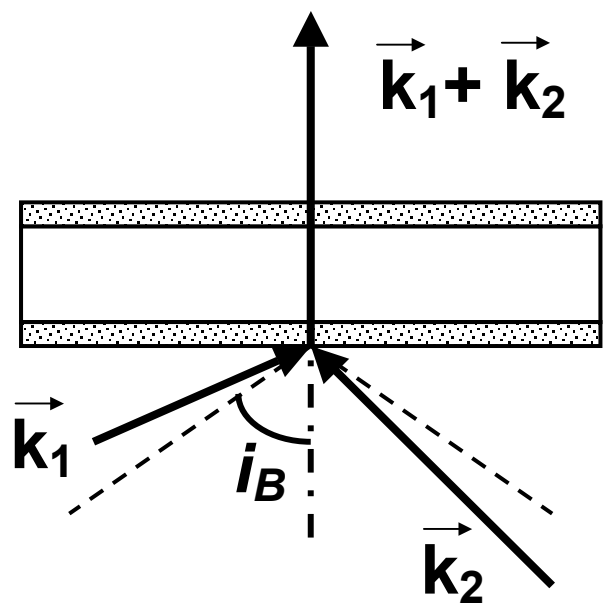

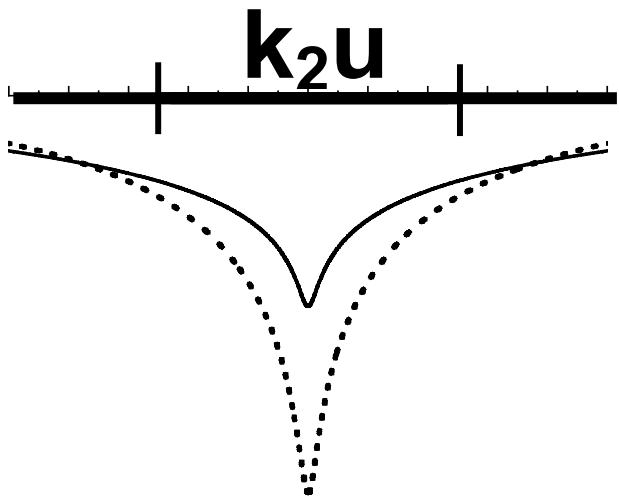

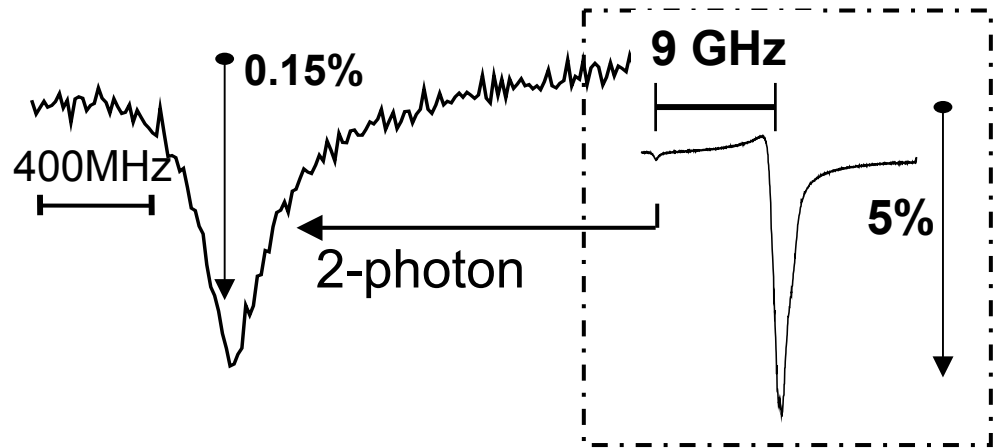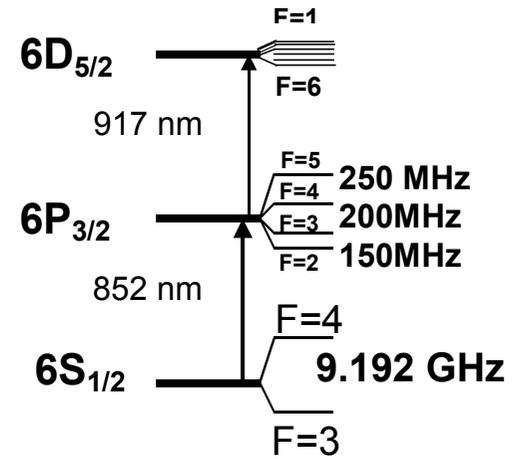